\documentclass[prd,preprintnumbers,twocolumn,eqsecnum,superscriptaddress,floatfix,nofootinbib,a4paper]{revtex4}
\usepackage{color}
\usepackage{calc}
\usepackage{amsmath,amssymb,graphicx}
\usepackage{amssymb,amsmath}
\usepackage{tensor}
\usepackage{bm}
\usepackage{times}
\usepackage{verbatim}
\usepackage[varg]{txfonts}
\usepackage[normalem]{ulem} 
\usepackage[colorlinks, pdfborder={0 0 0}]{hyperref}
\definecolor{LinkColor}{rgb}{0.75, 0, 0}
\definecolor{CiteColor}{rgb}{0, 0.5, 0.5}
\definecolor{UrlColor}{rgb}{0, 0, 0.75}
\hypersetup{linkcolor=LinkColor}
\hypersetup{citecolor=CiteColor}
\hypersetup{urlcolor=UrlColor}
\allowdisplaybreaks
\newcommand{\<}{\begin{equation}}
\newcommand{\?}{\end{equation}}

\newcommand{\cE}{\mathcal{E}}
\newcommand{\cF}{\mathcal{F}}
\newcommand{\cM}{\mathcal{M}}
\newcommand{\cR}{\mathcal{R}}

\newcommand{\R}{\mathbb{R}}
\newcommand{\N}{\mathbb{N}}

\newcommand{\fe}{\mathfrak{e}}
\newcommand{\ff}{\mathfrak{f}}

\newcommand{\inorm}[1]{\| #1 \|_\infty}
\newcommand{\inorml}[1]{\left\| #1 \right\|_\infty}
\newcommand{\tnorm}[1]{\| #1 \|_2}
\newcommand{\tip}[2]{\langle #1 , #2\rangle_2}
\newcommand{\pnorm}[1]{\| #1 \|_p}
\newcommand{\pnorml}[1]{\left\| #1 \right\|_p}
\definecolor{mgreen}{rgb}{0.1,0.8,0.25}

\begin{document}

\title{Estimating effective higher order terms in the post-Newtonian binding energy and gravitational-wave flux: Non-spinning compact binary inspiral}
\author{Shasvath J.~Kapadia}
\affiliation{Department of Physics, University of Arkansas, Fayetteville, Arkansas 72701, USA}
\affiliation{International Centre for Theoretical Sciences, Tata Institute of Fundamental Research, Bangalore 560012, India}
\author{Nathan~K.~Johnson-McDaniel}
\author{Parameswaran~Ajith}
\affiliation{International Centre for Theoretical Sciences, Tata Institute of Fundamental Research, Bangalore 560012, India}

\begin{abstract}
In the adiabatic post-Newtonian (PN) approximation, the phase evolution of gravitational waves (GWs) from inspiralling compact binaries in quasicircular orbits is computed by equating the change in binding energy with the GW flux. This \emph{energy balance} equation can be solved in different ways, which result in multiple \emph{approximants} of the PN waveforms. Due to the poor convergence of the PN expansion, these approximants tend to differ from each other during the late inspiral. Which of these approximants should be chosen as templates for detection and parameter estimation of GWs from inspiraling compact binaries is not obvious. In this paper, we present estimates of the \emph{effective} higher order (beyond the currently available 4PN and 3.5PN) non-spinning terms in the PN expansion of the binding energy and the GW flux that minimize the difference of multiple PN approximants (TaylorT1, TaylorT2, TaylorT4, TaylorF2) with effective one body waveforms calibrated to numerical relativity (EOBNR). We show that PN approximants constructed using the effective higher order terms show significantly better agreement (as compared to 3.5PN) with the inspiral part of the EOBNR. For non-spinning binaries with component masses $m_{1,2} \in [1.4 M_\odot, 15 M_\odot]$, most of the approximants have a match (faithfulness) of better than 99\% with both EOBNR and each other.
\end{abstract}
\preprint{LIGO-P1500166-v3}
\preprint{ICTS/2015/8}
\maketitle

\section{Introduction and summary}\label{sec:Introduction}
One of the biggest scientific enterprises of recent times, the quest for the direct detection of gravitational waves (GWs), is expected to achieve its first success in the near future. Some of the second-generation of interferometric GW detectors~\cite{0264-9381-27-8-084006,AdvVirgo:2009} will start operating later this year and are expected to achieve their design sensitivity over the next few years~\cite{Aasi:2013wya}. Estimates of the astrophysical rates of candidate GW sources, in particular coalescing compact binaries, predict that first detections are within the reach of these observatories~\cite{Abadie:2010cf}. 

GW signals from coalescing compact binaries, buried in the noisy detector data, are to be detected by cross-correlating the data with theoretical templates of expected signals. These theoretical GW templates are computed by solving the field equations of General Relativity for the two body systems. The two-body problem in General Relativity has no analytical exact solution. Hence the construction of GW templates either requires approximation techniques that help to tackle the problem analytically, or large-scale numerical computations for solving the problem exactly. In the early stages of the \emph{inspiral} of the compact binaries, where the orbit can be approximated as an adiabatic sequence of quasi-circular orbits, the GW templates can be computed using the post-Newtonian (PN) approximation to General Relativity~\cite{Blanchet:LivRev}. However, the modeling of the dynamics of the system and the GW signals from the late inspiral and merger stages requires large-scale numerical relativity simulations~\cite{Centrella:2010mx}. (See, e.g., the discussion in~\cite{Ohme2011} about the region in which the PN description of the inspiral is valid.)

For ``low-mass'' binaries (total mass $\lesssim 12 M_\odot$), the GW signal observed by ground-based interferometric detectors will consist almost entirely of the inspiral portion of the waveform~\cite{Ajith:2007xh,biops}, which could, in principle, be modeled accurately by the adiabatic PN approximation. In this approximation, the phase evolution of the binary's quasi-circular orbit (and hence the gravitational waveform) is computed by equating the loss of the orbital binding energy with the energy flux of the GWs. This \emph{energy balance} equation can be solved in different ways, which result in multiple \emph{approximants} of the PN waveforms. Due to the poor convergence of the PN expansion, these approximants tend to differ from each other during the late inspiral stage. Which of these approximants should be used as templates for detection and parameter estimation of GWs from inspiraling compact binaries is not obvious. Luckily, for the case of non-spinning binaries, most of the standard PN approximants are shown to be \emph{effectual}~\cite{DIS98} for the purpose of GW detection; however they are not \emph{faithful} to the actual signals for accurate estimation of the source parameters~\cite{biops}. For the case of highly spinning binaries, on the other hand, the currently available PN approximants fail to be even effectual for GW detection using advanced detectors~\cite{Nitz2013}.

Motivated by this issue in GW data analysis, we seek to estimate \emph{effective} higher order terms in the PN expansion of the binding energy and GW flux such that multiple, if not all, PN approximants have close agreement with a fiducial ``exact'' waveform family. We do so by fitting the PN approximants of an appropriate dynamical quantity [which we choose to be the evolution of the PN time $t(v)$ as a function of the PN expansion parameter $v$] with that computed from a fiducial exact waveform over a range of mass ratios. This fitting is done over the putative inspiral regime --- frequencies less than that of the Schwarzschild innermost stable circular orbit ($v = 1/\sqrt{6}$). As the fiducial exact waveform family, we choose the effective one body waveforms calibrated to numerical relativity (EOBNRv2)~\cite{Pan:2011gk}. We estimate two effective higher order terms --- or ``pseudo-PN'' (pPN)  terms --- in the binding energy (log-independent and log-dependent terms at 5pPN order), and five effective higher order terms in the GW flux (log-independent term at 4pPN, and log-independent and log-dependent terms at 4.5pPN, and 5pPN order). These are shown in Fig.~\ref{fig:pPN_eta_fit}. Currently we restrict ourselves to the case of non-spinning binaries. 

We show that multiple waveform approximants (TaylorT1, TaylorT2, TaylorF2) generated using the 5pPN accurate energy and flux functions show excellent agreement (faithfulness $\simeq 0.99 - 0.999$) with the fiducial exact waveform family (EOBNRv2) over the whole ``low-mass'' parameter space $m_{1,2} \in [1.4 M_\odot, 15 M_\odot]$ (see Figs.~\ref{fig:mismatch_EOB} and \ref{fig:contour_mismatch}). The TaylorT4 approximant shows very good agreement (faithfulness $\simeq 0.99$) over most, but not all, of the parameter space. For the majority of the cases, the faithfulness of the pPN approximants are significantly better than that at 3.5PN order (see Fig.~\ref{fig:mismatch_EOB}). Since the other adiabatic PN approximants are all basically variants of TaylorT1, TaylorT2, and TaylorT4, we expect these results to hold for other approximants also. These pPN coefficients can be readily applied for searches for GWs from non-spinning low-mass binaries. Work is ongoing to extend this method to the case of spinning binaries. 

We stress that the higher order terms that we estimate are \emph{effective} terms, in the sense that they capture the effects of a large number of actual higher order terms in the PN expansion. Hence they will be different from the \emph{actual} terms at a given PN order. This is in contrast with the work that has been done in the extreme mass-ratio limit, where one is able to determine true PN coefficients in the flux and binding energy (in addition to spin precession and tidal effects), since one can work to very high precision (up to thousands of digits) and at very large radii (up to $10^{70}$ times the Schwarzschild radius of the central black hole), allowing one to easily disentangle the individual PN coefficients, and to extract their analytical expressions (see \cite{Shah2014,Johnson-McDaniel2015} for application of these methods to the binding energy). We also note that there is related work by Huerta \emph{et al.}~\cite{Huerta2014}, who similarly fit for effective PN coefficients using EOB waveforms in the context of intermediate-mass-ratio inspirals. 

The rest of this paper is organized as follows: Section~\ref{sec:adiabatic_pn} briefly reviews the adiabatic energy-balance equations, as well as the computation of different PN approximants of the GW phase evolution. Sec.~\ref{sec:Intro-pPN} introduces the pseudo-PN coefficients, describes our method for determining them, and then evaluates their performance by computing mismatches between EOBNR waveforms and PN waveforms. The concluding section~\ref{sec:conclusions} summarizes the results and suggests possible avenues for future work. We give various ancillary technical results as appendices. Throughout this paper we use geometrized units: $G = c = 1$.

\section{Post-Newtonian waveforms in the adiabatic approximation}
\label{sec:adiabatic_pn}

A compact binary revolving about its center of mass radiates away orbital binding energy via the emission of GWs. This radiation in turn causes the separation $R$ between the component masses to shrink, and the orbit decays. For a large portion of the binary's inspiral, the rate of shrinkage of $R$ (or, equivalently, the rate of increase of angular speed) is negligible over the duration of an orbit. During this
\emph{adiabatic regime}, the rate of loss of binding energy $\mathcal{E}$ may be equated with the GW flux $\mathcal{F}$ radiated.\footnote{Actually, one can only equate the GW flux to the system's mechanical energy loss up to the Schott term. However, this term vanishes for exactly circular orbits and is thus small for quasicircular ones when one is in the adiabatic regime; see, e.g., the calculation of the Schott contribution in the EOB framework and associated discussion in~\cite{Bini2012}.} For a quasicircular binary, this relation alone results in a set of coupled ordinary differential equations for the orbital evolution, called the \emph{phasing formula}~\cite{DIS01}: 
\begin{equation}
\frac{dv}{dt} = -\frac{\mathcal{F}(v)}{\mathcal{E}'(v)}, \qquad\qquad \frac{d\varphi}{dt} = \frac{v^3}{M},
\label{energybalance}
\end{equation}
where $M$ is the total mass of the binary.
The orbital binding energy $\mathcal{E}(v)$ and the GW flux $\mathcal{F}(v)$ can be computed as PN expansions in terms of a gauge-invariant velocity parameter $v$~(see, e.g.,~\cite{Blanchet:LivRev} for a review). These equations, which describe the time evolution of the orbital phase $\varphi$ and the velocity parameter $v$, may be written in integral form as:
\begin{subequations}
\label{t_phi_v}
\begin{align}
t(v) &= t_{\mathrm{ref}} + \int^{v_{\mathrm{ref}}}_{v}\frac{\mathcal{E}'(\bar{v})}{\mathcal{F}(\bar{v})}d\bar{v}, \\
\varphi(v) &= \varphi_{\mathrm{ref}} + \frac{1}{M}\int^{v_{\mathrm{ref}}}_{v}\bar{v}^3\frac{\mathcal{E}'(\bar{v})}{\mathcal{F}(\bar{v})}d\bar{v}.
\end{align}
\end{subequations}
These integral expressions describe the time and phase evolution of the orbit as a function of the expansion parameter $v$. Here, $v_\mathrm{ref}$ is a reference value of $v$ while $t_{\mathrm{ref}}$ and $\varphi_{\mathrm{ref}}$ represent the time and phase of the orbit at $v = v_\mathrm{ref}$.

The phasing formula can be solved in a number of different methods which are perturbatively equivalent, in the sense that they all are accurate to a given PN order. They can be generally classified into three approaches: 

\begin{enumerate} 
\item Compute the energy and flux functions in Eq.~(\ref{energybalance}) up to a given PN order, evaluate the ratio $\mathcal{F}(v)/\mathcal{E}'(v)$ numerically, and solve the ordinary differential equations using an appropriate numerical method. The resulting time-domain approximant is
known as ``TaylorT1''~\cite{DIS01}. One can also compute an equivalent frequency domain approximant, ``TaylorF1'' making use of the stationary phase approximation~\cite{DIS01}.
\item Re-expand the ratio $ \mathcal{F}(v) / \mathcal{E}'(v)$ in Eq.~(\ref{energybalance}) as a power series and truncate it at the respective PN order. The resulting time-domain approximant obtained by solving the ordinary differential equations numerically is called ``TaylorT4''~\cite{Boyle2007}. Frequency domain equivalents of TaylorT4, such as ``TaylorF4'' and ``TaylorR2F4'' can be computed via the stationary phase approximation~\cite{Nitz2013}. 
\item Re-expand the ratio $\mathcal{E}'(v) / \mathcal{F}(v)$ in the integral form of the phasing formula~\eqref{t_phi_v} as a power series and truncate it at the respective PN order. This allows us to evaluate the integrals analytically resulting in a parametric representation in terms of $t(v)$ and $\varphi(v)$, from which one can obtain $\varphi(t)$. This is known as the 
``TaylorT2'' approximant~\cite{DIS01}. Other time-domain approximants making use of this re-expansion include ``TaylorT3''~\cite{DIS01} and ``TaylorT5''~\cite{Ajith:2011ec}. Based on this re-expansion, one can also compute a frequency domain approximant, ``TaylorF2,'' making use of the stationary phase approximation~\cite{DIS01}.
\end{enumerate} 

The energy function $\mathcal{E}(v)$ and the flux function $\mathcal{F}(v)$ are known only to a finite PN order: For the case of non-spinning binaries, $\mathcal{E}(v)$ is known to 4PN order \cite{Jaranowski2013, Damour2014}, and $\mathcal{F}(v)$ to 3.5PN order~\cite{Blanchet2002, Blanchet2004}. (See also~\cite{Blanchet:LivRev} for a review of these computations.) The approximants therefore increasingly diverge from each other as the inspiral progresses and $v$ gradually increases to a considerable fraction of the speed of light. This divergence is not acute during early-inspiral ($v \ll 1$), but can become considerable during late inspiral \cite{Boyle2007, Nitz2013}. An obvious conundrum that arises is the choice of approximant to make to better model the ``true'' GW waveforms of Nature during late inspiral, leading to merger. Waveforms that more accurately represent GWs from inspiraling compact binaries exist, such as those produced by large-scale numerical-relativity (NR) computations. In principle one could use these waveforms and do away with PN waveforms.  However, this is not feasible in practice, given the enormous computational cost of NR computations. EOBNR waveforms are an excellent substitute for NR waveforms (at least in the non-spinning case). There is unfortunately still a non-trivial computational cost associated with producing EOBNR waveforms, in view of the fact that millions of them will be required to construct a template bank suitable for GW observations.\footnote{This issue is partly solved by the development of surrogate models of EOBNR waveforms~\cite{Purrer:2014fza} making use of reduced-order modeling techniques~\cite{Field:2011mf,Field:2012if,Field:2013cfa}. However, the construction of these reduced-order models requires the generation of tens of thousands of EOBNR waveforms, and then finding an orthonormal basis for them. This also incurs a significant computational and memory cost.} In the next section, we propose to introduce effective higher order terms in the PN expansion of the binding energy and GW flux as a method to overcome this. 

\section{Introducing pseudo-PN terms in the energy and flux}
\label{sec:Intro-pPN}
Circumventing the onerous task of computing higher order PN terms analytically (which would still not have necessarily improved the accuracy of the approximants), we introduce a single set of higher order {\it effective} PN coefficients in the energy and flux of the binary to simultaneously improve the agreement (as quantified by \emph{faithfulness}~\cite{DIS98}) between waveforms produced by multiple approximants, and their EOBNR counterparts.  

The PN formalism expresses the binding energy and flux as expansions in powers of $v$. So far, binding energy PN coefficients up to eighth order in $v$ beyond the Newtonian order (4PN) have been determined; furthermore, for non-spinning binaries, coefficients at half-integer PN orders are known to vanish through $4.5$PN, but start to be nonzero at $5.5$PN~\cite{Shah2014,Bini:2013rfa,Blanchet:2013txa}. On the other hand, PN flux coefficients have been computed up to 3.5PN, and unlike for the binding energy, there are nonzero coefficients at all half integer PN orders starting from $1.5$PN. 

We propose determining the pPN coefficients for $\mathcal{E}(v)$ and $\mathcal{F}(v)$ up to 5pPN order, starting from the lowest order at which PN coefficients have thus far not been determined. We assume the following ansatz for our pPN coefficients, guided by the form of known PN terms, some of which include quantities proportional to $\ln v$ (known from the test-mass limit of the energy flux---see, e.g.,~\cite{Fujita2012}---and the first-order self force results for the binding energy given in, e.g.,~\cite{Bini:2013rfa}):     
\begin{subequations}
\begin{align}\label{PN-EnergyFlux}
\mathcal{E}_{5\text{pPN}}(v) &= -\frac{1}{2}M\eta v^2 \,   \left[\sum_{k = 0}^{8}\mathcal{E}_kv^k +  \left(E_{\mathrm{10}} + E_{\mathrm{10}}^\mathrm{L}\ln v\right)v^{10} \right], \\
\mathcal{F}_{5\text{pPN}}(v) &= \frac{32}{5} \eta^2 v^{10} \, \left[\sum_{k = 0}^{7}\mathcal{F}_kv^k + \sum_{k = 8}^{10}\left(F_k + F_{k}^{\mathrm{L}}\ln v\right)v^k\right],
\end{align}
\end{subequations}
Here, $\eta := m_1 m_2/M^2$ is the symmetric mass ratio of the binary, where $m_1$ and $m_2$ are the individual masses, $\mathcal{E}_k$, $\mathcal{F}_k$ are the known PN coefficients, and $E_{10}$, $E^{\mathrm{L}}_{10}$, $F_k$, $F_{k}^{\mathrm{L}}$ are the pPN and pPN-log terms to be determined by calibrating PN quantities such as $t(v)$ or $\varphi(v)$ given in Eqs.~(\ref{t_phi_v}) to their EOBNR counterparts. Just like ordinary PN coefficients, we expect pPN terms to vary smoothly with the symmetric mass ratio $\eta$, ideally as low order polynomials in $\eta$. 

\subsection{Determining the pseudo-PN coefficients}
\label{sec:pPN_coeffs}

\begin{figure*}[tbh]
\includegraphics[width=0.95\textwidth]{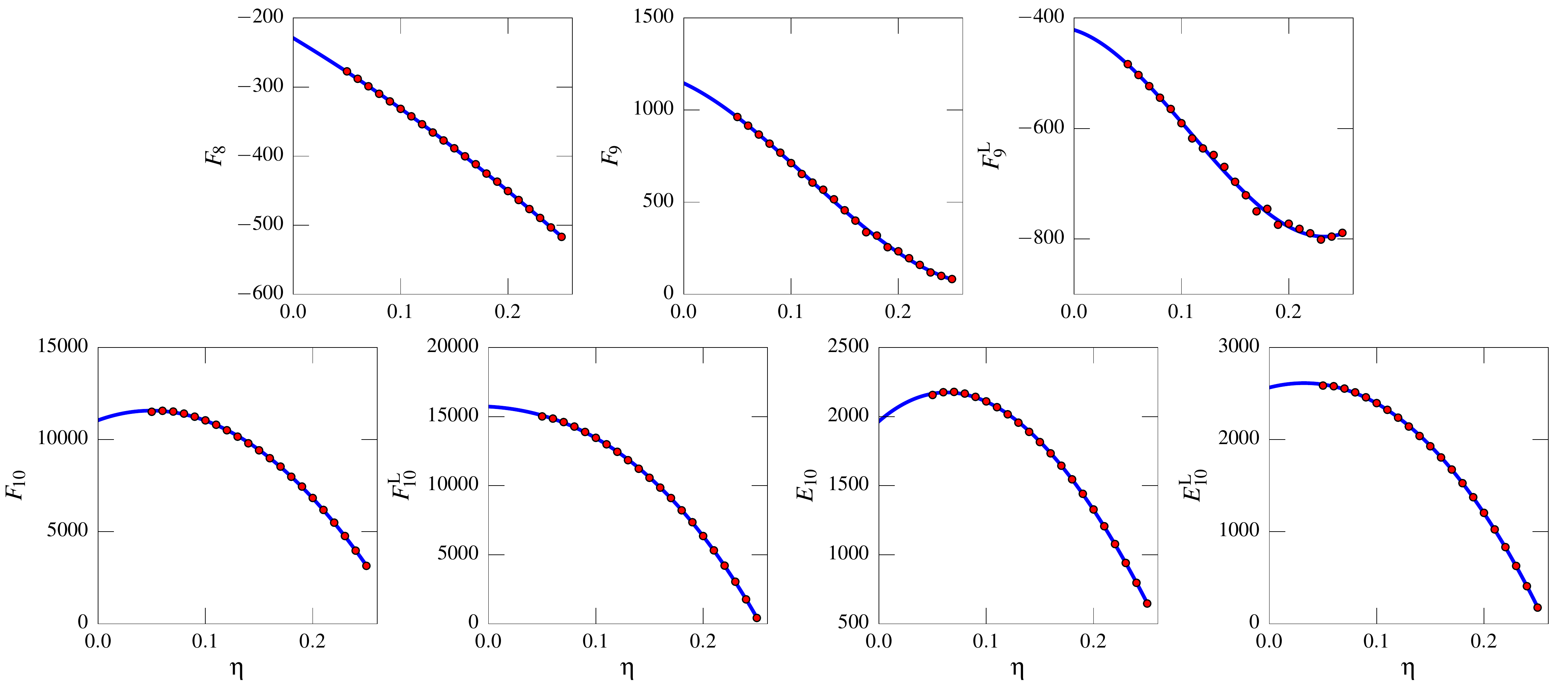}
\caption{The estimated pseudo-PN coefficients (red points) plotted against the symmetric mass ratio $\eta$ of the binary, along with the polynomial fits to these data points given by Eqs.~\eqref{pPN_eta}. (We do not
plot $F_8^\mathrm{L}$ here, since we set it to the test-particle value for all $\eta$.)}
\label{fig:pPN_eta_fit}
\end{figure*}

As mentioned earlier, waveforms generated by large scale NR computations are the fiducial waveforms of choice, to which we could calibrate our PN waveforms and determine the pPN coefficients. However, NR waveforms spanning the long inspiral are computationally expensive, and are thus not yet available for binaries with arbitrary mass ratios. In order to lift this restriction, we use instead the EOBNRv2 waveform model~\cite{Pan:2011gk} as implemented in the \textsc{LALSimulation} package, part of the \textsc{LALSuite} software library~\cite{LAL}.

We broadly considered two ways by which we could estimate our pPN terms. In the spirit of TaylorT1, the first method we considered attempts to extract pPN coefficients by fitting the ratio $-{\mathcal{F}(v)}/{\mathcal{E}'(v)}$ (keeping the energy and flux to 5pPN order) to the velocity derivative $dv/dt$ computed using EOBNR, over a range of $v$ spanning the late inspiral, say, $v \in [0.2, v_\text{ISCO}]$, where $v_\text{ISCO} = 6^{-1/2}$ is the velocity at the innermost stable circular orbit (ISCO) for a test particle orbiting a Schwarzschild black hole.

As expected, the agreement between TaylorT1 and EOBNR improves noticeably. For example, the faithfulness for mass combination $1.4M_{\odot}-1.4M_{\odot}$ increased from about 0.95 at 3.5PN to about 0.999 at 5pPN. However, the pPN coefficients so determined were significantly larger (by three to four orders of magnitude) than the average size of PN coefficients.\footnote{One can gauge the order of magnitude of PN coefficients by looking at the test particle limit, where these coefficients are known to high orders. One finds that the $n$PN coefficient increases in size roughly as $3^{n/2}$, due to the divergence at the light ring (see, e.g., Fig.~3 in~\cite{Johnson-McDaniel2015}), so that at the pPN orders we are considering, we expect flux coefficients on the order of $10^2$ to $10^3$ and binding energy coefficients on the order of $10^1$ to $10^2$; see the explicit expressions for the test particle flux in~\cite{Fujita2012} and the general form of the test particle binding energy in, e.g., Eq. (3.3) of~\cite{DIS98}.} Other approximants computed using these pPN terms exhibited poor faithfulness with the EOBNR waveforms. Including the known terms from the test-particle limit in the energy and flux (to $22$PN, using the exact energy and the flux from~\cite{Fujita2012}) did not tame these pPN coefficients -- they still remained undesirably large.    

The second method, which, after some tuning, yielded promising results, computes $t(v)$ from Eq.~(\ref{t_phi_v}) using TaylorT2, up to 5pPN order. The coefficients of $t(v)$ at orders beyond 3.5PN are functions of the pPN quantities $\{E_{10}$, $E_\mathrm{10}^\mathrm{L}$, $F_{8,9,10}, F^\mathrm{L}_{8,9,10}\}$. (We chose this number of coefficients since it is the smallest number that gave us the level of agreement we desired in the final matches, and also includes all the expected terms in both the energy and flux at a given PN order; we have not experimented extensively with adding further coefficients.) Instead of fitting the $t(v)$ we obtain from these coefficients to its corresponding EOBNR analogue, and thus estimating the energy and flux pPN coefficients directly, we choose instead to define a functional form for the TaylorT2 pPN terms as follows: 

\begin{equation}
\label{tt2}
t_\mathrm{5pPN}(v) = \frac{5M}{256 \eta v^8} \, \left(\sum_{k = 0}^{7}t_kv^k+\sum_{k = 8}^{10}\Theta_k v^k\right) 
\end{equation}
where
\begin{subequations}
\label{tt2_ppn}
\begin{align}
\Theta_8 &= \theta^\mathrm{L}_{\mathrm{8}}\ln v + \theta^{\mathrm{L}2}_{\mathrm{8}}\ln^2 v,\label{tt2_ppn1}\\
\Theta_k &= \theta_k + \theta_{k}^\mathrm{L}\ln v, \quad k \in \{9, 10\}  \label{tt2_ppn2}. 
\end{align}
\end{subequations}
This ansatz closely parallels the actual TaylorT2 expansion of $t(v)$ (see~\cite{biops} for the expression through $3.5$PN and~\cite{Varma2013} for the expression to higher orders in the extreme mass-ratio limit); as a result, to the orders we consider, the $\theta$ coefficients are all {\it linear} in terms of the pPN coefficients of the energy and flux. The higher order terms in the $t(v)$ expansion can be written in terms of the higher order PN coefficients in the energy and flux (to six significant digits) as: 
\begin{subequations}\label{constraints}
\begin{align}
\theta_\mathrm{8}^\mathrm{L} &= 2730.98 - 1015.30\eta   - 225.890\eta^2 - 45.0201\eta^3\nonumber\\
&\quad - 44.0104\eta^4  + 8F_8,\\
\theta_\mathrm{8}^{\mathrm{L}2} &= -386.268 - 1564.24\eta + 4F_{8}^\mathrm{L}, \label{theta8L2}\\
\theta_{9} &= -12572.5 + 15468.4\eta + 10627.8\eta^2 + 2328.76\eta^3+ 8F_9\nonumber\\
&\quad   - 8F_{\mathrm{9}}^\mathrm{L},\\
\theta_\mathrm{9}^\mathrm{L} &= 3278.27 + 8F_{\mathrm{9}}^\mathrm{L},  \\
\theta_{10} &=  15242.5 - 5463.77\eta + 3469.49\eta^2+ 863.828\eta^3 \nonumber\\ 
&\quad - 402.824\eta^4 - 33.9446\eta^5 + (23.6905 +22.6667\eta) F_8 \nonumber\\
&\quad  - (11.8452 + 11.3333\eta)F_\mathrm{8}^\mathrm{L} - 24E_{10} + 10E_\mathrm{10}^\mathrm{L} + 4F_{10}   \nonumber \\
&\quad  - 2F_\mathrm{10}^\mathrm{L} \label{constr1},\\
\theta_\mathrm{10}^\mathrm{L} &= -1951.32 -6083.57\eta- 4606.30\eta^2 + (23.6905 \nonumber \\
&\quad + 22.6667\eta)F_{\mathrm{8}}^\mathrm{L} - 24E_{\mathrm{10}}^\mathrm{L} + 4F_{\mathrm{10}}^\mathrm{L}.
\label{constr2}
\end{align}
\end{subequations}

\begin{figure*}[tbh]
\includegraphics[width=0.85\textwidth]{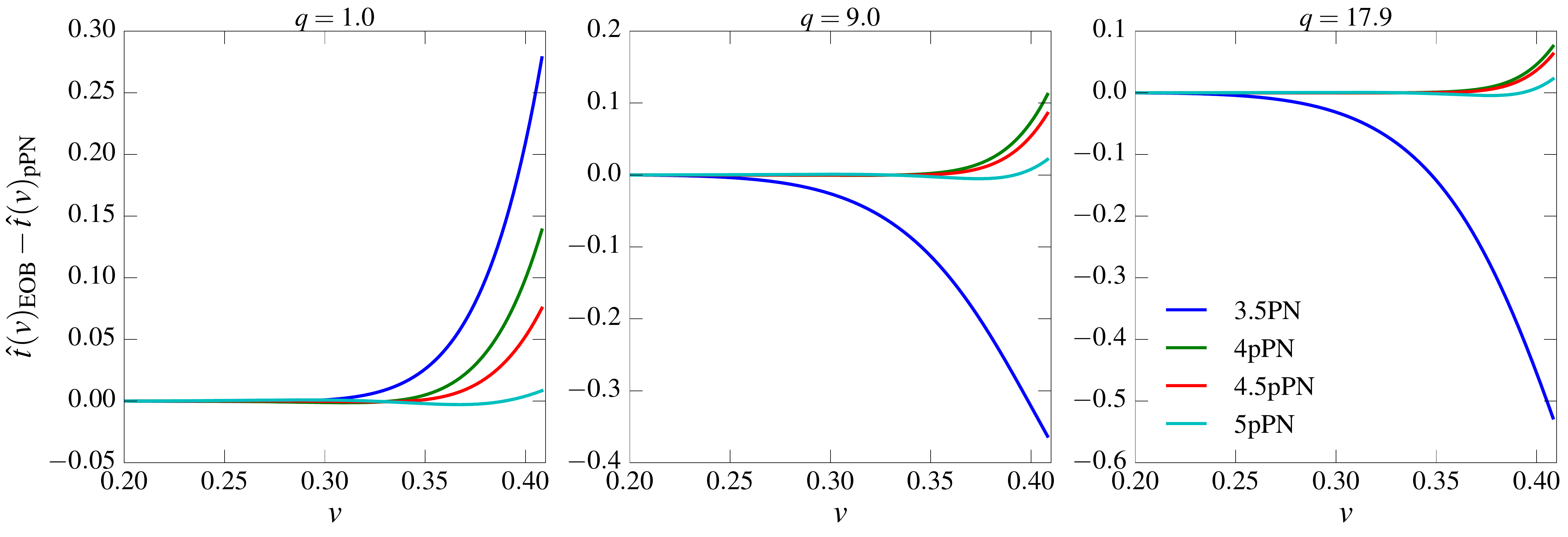}
\caption{The difference between $t(v)$ as computed via EOBNR and TaylorT2 (both rescaled by the Newtonian value of $t$), at orders from 3.5PN to 5pPN, for three different mass ratios. The improvement in the agreement  between EOBNR and TaylorT2 (due to the addition of pPN terms) becomes more dramatic with increasing mass ratio.} 
\label{fig:residuals}
\end{figure*}

The next step involves fitting our ansatz for the $t_\mathrm{5pPN}(v)$ given in Eq.~\eqref{tt2} to its EOBNRv2 counterpart $t_\mathrm{EOB}(v)$ for a range of mass ratios. We compute $v = (M\omega)^{1/3}$ from the dominant quadrupole $(\ell = m = 2)$ mode $\mathfrak{h}_{22}$ of the EOBNRv2 waveforms, where the orbital frequency $\omega$ is computed in the following way: 
\begin{equation}
\label{omega_comp}
\omega = \frac{1}{2} \, \frac{d\varphi_{22}}{dt}, \qquad\varphi_{22} = \arg(\mathfrak{h}_{22}).
\end{equation}
Here we compute the derivative using second-order centered finite differencing of the numerical data. We then fit the resulting data for $t_\mathrm{EOB}(v)$ to the analytical expression $t_\mathrm{5pPN}(v)$ using a least-squared minimization algorithm, to obtain numerical values of the pPN coefficients $\theta_k, \theta_k^\mathrm{L}$ and $\theta_k^{\mathrm{L}2}$ for different mass ratios. Here we use a least-squares fit (i.e., minimize the $L^2$ norm of the difference of the functions) not only because it is a standard fitting procedure, but also because we are primarily interested in improving the matches between the functions, which are fairly closely related to the $L^2$ norm.\footnote{Indeed, by Plancherel's theorem, in the case of white noise, maximizing the match is equivalent to minimizing the $L^2$ norm of the difference between the functions, since in this case the match is just the $L^2$ inner product between the ($L^2$) normalized functions. In fact, under appropriate simplifying assumptions, one can obtain an explicit lower bound on the match between the two waveforms in terms of the $L^2$ norm of the difference of their phases, given in Appendix~\ref{L2_bounds}.} Additionally, as discussed in Appendix~\ref{phi_bound}, the pPN coefficients obtained from a least square fit in $t(v)$ automatically improve the least square residual of $\varphi(v)$. In performing these fits, we have the freedom to set the EOB and pPN $t(v)$s to be equal to each other at a reference value of $v$ by adding a constant: We choose to do this at $v = 0.2$.

While an obvious approach would be to fit for all these coefficients over the entire range of $v$ values we are considering, we found that we obtained better results (possibly closer to the actual PN coefficients) if we used a more involved method. In this method, we realize that the lowest-order coefficients we consider (i.e., $F_8$ and $F_8^\mathrm{L}$ at $4$pPN) will be dominant at small $v$, so it makes sense to only fit for those coefficients over a restricted range. However, we do not know this range \emph{a priori}, so we let the upper limit of the interval over which we fit (the transition velocity $v_{8}^t$) vary and minimize the residual over the entire range of $v$ we consider $[v_\text{min}, v_\text{max}]$, to determine $v_{8}^t$ (which we take to be at least slightly larger than $v_\text{min}$). We then subtract off the contribution of the $4$pPN terms and move on to the $4.5$pPN terms, where we apply the same method, fitting over the interval $[v_\text{min}, v_{9}^t]$, requiring that $v_{8}^t < v_{9}^t < v_\text{max}$. For the $5$pPN terms, we fit over the entire interval, since we want to improve the overall agreement as much as possible, and are not adding on any more terms. This iterative method was inspired by the iterative method for obtaining the PN coefficients in the linear-in-$\eta$ portion of the binding energy from a high-precision self-force calculation used in~\cite{Johnson-McDaniel2015}. The fits we quote are done using $v_\text{min} = 0.2$ and $v_\text{max} = v_\text{ISCO}$.

The fit-and-minimize-over-transition-velocities method we just described yields some unwanted jumpiness in the pPN coefficients, the largest components of which can be attributed to a similar jumpiness in the dependence of the transition velocities on $\eta$. In order to reduce this jumpiness, we thus fix the values of the transition velocities to $v_8^t = v_9^t = 0.355$, which are very close to the values for these quantities returned by the above procedure for an equal-mass system (and a number of other mass ratios; indeed, almost all the $v_9^t$s given by the minimization are extremely close to $0.355$). We find that these fixed transition velocities give final residuals at $5$pPN that are quite similar to the residuals obtained from the fits that allow the transition velocities to vary, while reducing much of the unwanted jumpiness in the pPN coefficients.

We also ended up not including a variable coefficient of $\ln^2(v)$ at 4pPN in the fit, since when we did so, we found that the values assigned to the correction to the test mass value in the $4$pPN $\ln(v)$ term in the energy flux by the fit were small and quite jumpy. We thus fixed the coefficient of $\ln^2(v)$ at 4pPN by taking the $4$pPN $\ln(v)$ term to just have its test-mass value for all $\eta$ and then went back and performed the fit-and-minimize-over-transition-velocities procedure again to arrive at the final transition velocity values quoted above.

We display the residuals we obtain with this method as a function of $v$ in Fig.~\ref{fig:residuals}, which plots the difference between $t$ as computed via EOBNR and the (p)PN $t(v)$ values (both scaled by the Newtonian value of $t$) for (p)PN orders from 3.5PN to 5pPN. The plot shows that the residuals get progressively smaller with increasing PN order beyond 3.5PN, a clear indication that the pPN coefficients systematically improve the agreement between TaylorT2 and EOBNR. This improvement in the residuals should thus translate into an improvement in matches, as discussed above.

Having determined the pPN coefficients $\theta_8^\mathrm{L}, \theta_8^{\mathrm{L}2}, \theta_9, \theta_9^\mathrm{L}, \theta_{10}, \theta_{10}^\mathrm{L}$, we can extract the energy and flux pPN coefficients via Eqs.~\eqref{constraints}. It is obvious, upon inspecting these equations, that the flux coefficients at 4 and 4.5pPN (viz., $F_8, F_8^\mathrm{L}, F_9, F_9^\mathrm{L}$) are uniquely determined. However, the coefficients $E_{10}, E_\mathrm{10}^\mathrm{L}, F_{10}, F_\mathrm{10}^\mathrm{L}$ are not fixed, and may in fact take any value provided they satisfy their constraint equations~\eqref{constr1} and \eqref{constr2}. Exploiting this freedom in the 5pPN energy and flux parameters, we endeavor to set them in a way so as to improve the agreement between the TaylorT1 and EOBNR waveforms. To that end, we fit the 5pPN flux-to-energy ratio ${\mathcal{F}(v)}/{\mathcal{E}'(v)}$ to EOBNR's $dv/dt$, after fixing the pPN coefficients at 4 and 4.5pPN to those determined via the method described above. We use the constraint equations Eqs.~\eqref{constr1} and \eqref{constr2} to write the 5pPN flux coefficients as a function of the 5pPN energy coefficients, and then perform a two-parameter minimization over the energy coefficients.

We compute the pPN terms in the energy and flux function making use of a set of 21 EOBNRv2 waveforms,  with symmetric mass ratio values evenly spanning the interval $\eta \in [0.05, 0.25]$, as shown in Fig.~\ref{fig:pPN_eta_fit}. Fitting low order polynomials in $\eta$ to the numerically computed pPN coefficients (and including the test-particle value to which we set $F_8^\mathrm{L}$ for all $\eta$, for completeness), we write these pPN coefficients (to six significant digits) as: 
\begin{equation}
\label{pPN_eta}
\begin{split}
F_{8} &= -229.100 - 934.582\eta - 861.481\eta^2, \\
F_8^\mathrm{L} &= 52.7431, \\
F_{9} &= 1146.18 - 2743.15\eta - 22150.5\eta^2 + 64309.3\eta^3, \\
F_9^\mathrm{L} &= -421.553 - 595.925\eta - 15568.9\eta^2 + 48222.8\eta^3,    \\ 
F_{10} &= 11051.9 + 21160.8\eta - 215292\eta^2 + 18123.0\eta^3,   \\
F_{10}^\mathrm{L} &=  15725.6 - 4275.47\eta -157235\eta^2 -279859\eta^3, \\
E_{10} &= 1966.08 + 6752.30\eta - 56757.4\eta^2 + 34764.9\eta^3, \\
E_{10}^\mathrm{L} &= 2565.84 + 2946.63\eta - 44485.0\eta^2 - 21789.4\eta^3.
\end{split}
\end{equation}
We plot these fits, along with the original pPN coefficients at discrete $\eta$ in Fig.~\ref{fig:pPN_eta_fit}. We see that these coefficients are generally of a reasonable size (and the original $\Theta$ coefficients---which we do not show explicitly---are generally of similar size to the test-particle TaylorT2 coefficients at these orders, given in~\cite{Varma2013}), except for the binding energy coefficients, which are a bit large. 

\paragraph*{Numerical methods employed:} In the first part of the estimation of the pPN coefficients, where we fit the EOBNR $t(v)$ using a TaylorT2-like ansatz, we minimize the residuals over the transition velocities using the Nelder-Mead downhill simplex algorithm implemented in \textsc{Scipy}'s~\cite{scipy} \textsc{optimize.minimize} function, and determine the pPN coefficients at each order using the Levenberg-Marquardt non-linear least square algorithm implemented in \textsc{Scipy}'s \textsc{optimize.curve\_fit} function. In the second part, where we determine the energy and flux pPN coefficients at 5pPN by minimizing the difference of a TaylorT1
expression for $dv/dt$ with that computed from EOBNR, we compute the EOBNR
$dv/dt$ using second-order centered finite differencing. In order to
alleviate the modulations in our numerically computed EOBNR $dv/dt$
(likely coming from the residual eccentricity in the EOBNR waveform
due to difficulty in setting exactly quasi-circular initial
conditions) we have generated all the EOBNR waveforms we consider
starting from a velocity of $v = 0.15$, while we only fit for $v \geq
0.2$. Furthermore, we smoothen the numerically computed $v(t)$ by
employing \textsc{Scikit-Learn}'s~\cite{scikit-learn-url,scikit-learn}
implementation of a machine learning algorithm called isotonic
regression.

\subsection{Testing the performance of pPN coefficients}

\begin{figure*}[tbh]
\includegraphics[width=0.98\textwidth]{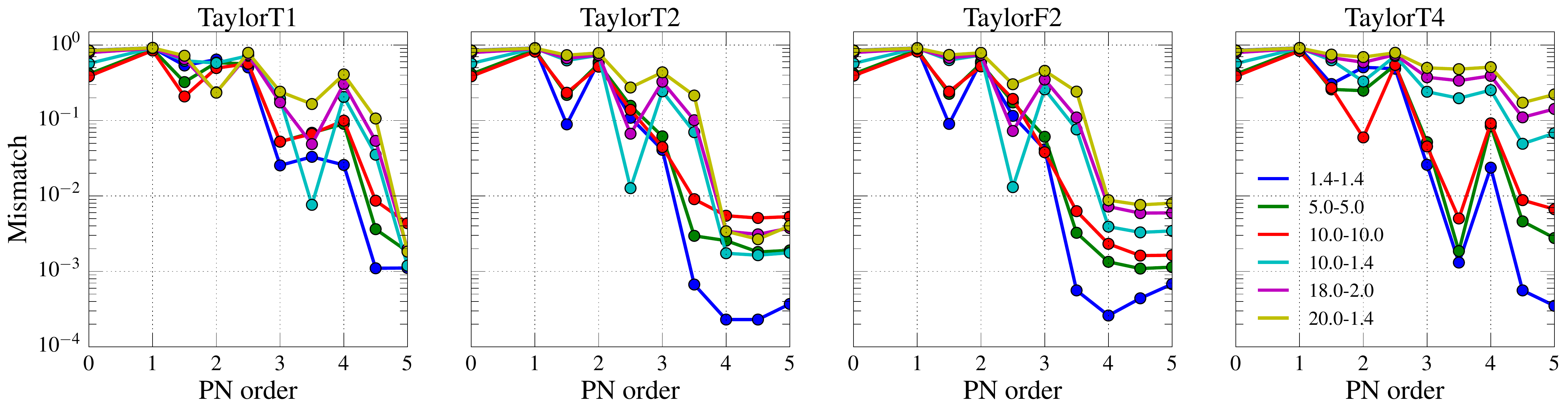}
\caption{Mismatch with EOBNR as a function PN order, for approximants TaylorT1, TaylorT2, TaylorT4, and TaylorF2. The legend shows the masses in $M_{\odot}$ of the binaries considered. For most of the approximants and mass combinations considered, the mismatch at 5pPN is not only better than the mismatch at 3.5PN, but is also lower than $10^{-2}$.} 
\label{fig:mismatch_EOB}
\end{figure*}

\begin{figure*}[tbh]
\includegraphics[width=0.98\textwidth]{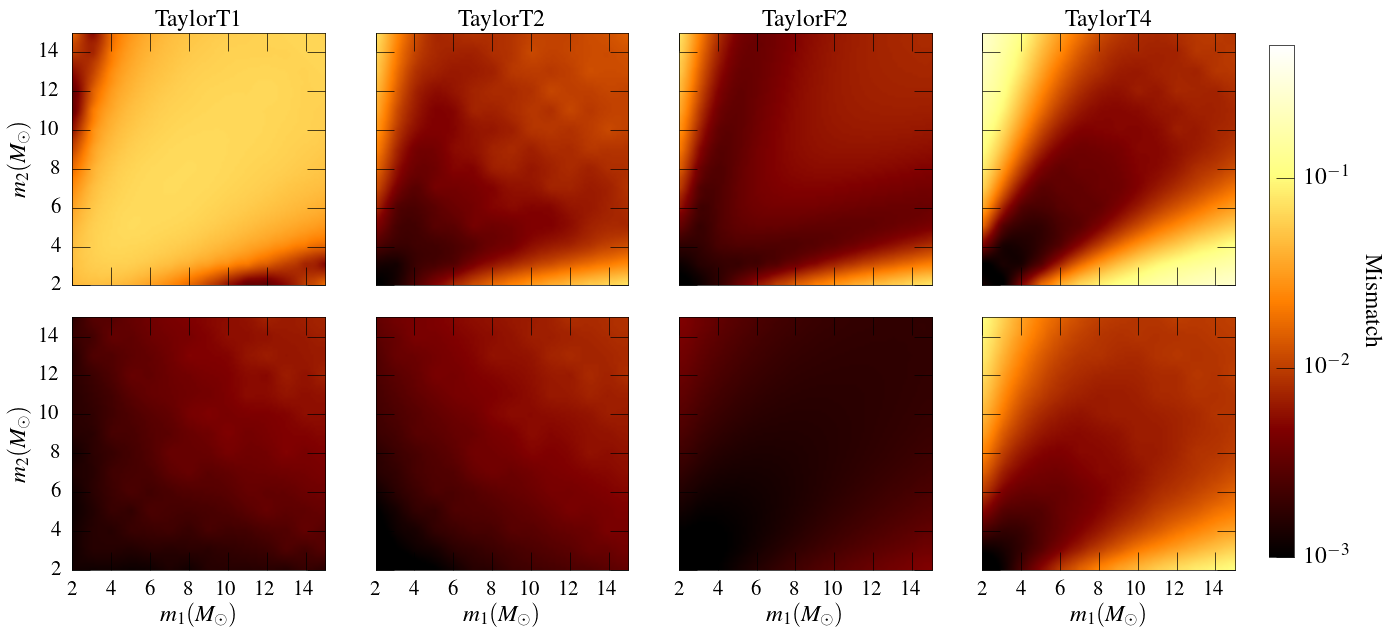}
\caption{Contour plots of the mismatch of different approximants with EOBNR over the mass range $m_{1,2} \in [2 M_{\odot}, 15 M_{\odot}]$. The top panels correspond to 3.5PN order and the bottom panel to 5pPN order. There is a consistent improvement in the mismatches at 5pPN, as compared to 3.5PN, for TaylorT1, TaylorT2, and TaylorF2, over the entire low-mass region examined here. The mismatches for TaylorT4 improve mostly for high-mass ratios; nevertheless, for most the low-mass region, the mismatches are near or below $10^{-2}$.}
\label{fig:contour_mismatch}
\end{figure*}

To evaluate the performance of our method of obtaining pPN coefficients in the energy and flux functions, we compute the \emph{mismatch} ($1 - \text{faithfulness}$) between the different approximants computed using these pPN coefficients and the corresponding EOBNR waveform. Following the discussion in Sec.~\ref{sec:adiabatic_pn}, we consider three approximants --- TaylorT1, TaylorT4, TaylorT2 --- which correspond to the three different ways of treating the ratio of the energy and flux functions appearing in the phasing formula. In addition, we also consider the frequency-domain TaylorF2 approximant. Although the behavior of the TaylorF2 approximant is expected to be very similar to that of TaylorT2, we explicitly consider the former because it is the approximant that is most widely used in GW data analysis.\footnote{Note that TaylorF2 is based on the stationary phase approximation, which obtains additional corrections starting at $5$PN, as discussed in~\cite{Varma2013}, which we do not include here.} We indeed see that the behavior of this approximant is very similar to that of TaylorT2. 

We evaluate the mismatches over the frequency range $[10~\mathrm{Hz},~f_\mathrm{ISCO}]$, where $10$~Hz corresponds to the lower cut-off of Advanced LIGO's expected frequency band, and $f_\mathrm{ISCO}$ is the dominant quadrupole mode GW frequency associated with the ISCO in the Schwarzschild geometry. The mismatches are computed assuming the ``high-power, zero-detuning'' noise power spectral density of Advanced LIGO~\cite{adligo-psd}. We consider a set of component masses $m_1-m_2$ consisting of standard fiducial systems $1.4 M_\odot - 1.4 M_\odot, 10 M_\odot - 1.4 M_\odot, 10 M_\odot - 10 M_\odot$, as well as three others, to better sample the space of total masses and mass ratios. (Note that the lower bound of the frequency band over which the mismatches are computed is smaller than the lower bound, $v=0.2$, of the range of $v$ over which the fits were conducted to compute the pPN coefficients for all the mass combinations we considered. The smallest $v$ that is in band is $v \simeq 0.08$ for the $1.4 M_\odot - 1.4 M_\odot$ system.) For each of these systems, we compute the mismatch as a function of the (pseudo) PN order, starting from 0PN and working our way up to 5pPN.\footnote{For the sake of consistency, we set the amplitude order equal to the PN order whenever possible, otherwise defaulting to the maximum amplitude order available in \textsc{LALSimulation}~\cite{LAL}, which is $3$PN for everything except TaylorF2, which only uses the Newtonian amplitude.} The results are summarized in Fig.~\ref{fig:mismatch_EOB}. We see that the pPN terms reduce the mismatch of all the approximants with EOBNR for almost all the cases considered here; in most cases the mismatches have been reduced to $< 10^{-2}$.

The only instance where the pPN terms have slightly worsened the mismatch is in the case of TaylorT4 approximant for comparable-mass binaries with large masses ($q \simeq 1,~ M > 10 M_\odot$). Given that none of the pPN coefficients were determined by improving the agreement between TaylorT4 and EOBNR, it is not surprising that the pPN coefficients are relatively less effective at improving their agreement. [In fact, TaylorT4 and TaylorT2 are, in some sense, maximally different from each other, since the former is based on the re-expansion of $\mathcal{F}(v)/\mathcal{E}(v)$, while the latter is based on the re-expansion of its inverse.] However, it is worthwhile to notice that the pPN coefficients still improve the matches at high mass ratios. Furthermore, for comparable mass systems, the mismatch at 5pPN, though marginally worse than at 3.5PN, still remains below $10^{-2}$.

Having looked at a set of discrete mass combinations, we further test the performance of our pPN terms by evaluating the mismatches over a continuous two-dimensional region of the $m_1-m_2$ parameter space, with $m_{1,2} \in [2 M_{\odot}, 15 M_{\odot}]$. Comparing the mismatches at 3.5PN with those at 5pPN (Fig.~\ref{fig:contour_mismatch}), we find, for three of the four approximants considered, a consistent reduction in mismatches at 5pPN (as compared to mismatches at 3.5PN) over the entire $m_1-m_2$ region displayed. For TaylorT4, the improvement in the mismatches occurs mostly for higher mass-ratio systems, while it does not become worse than $\sim 10^{-2}$ at comparable masses.  

\begin{figure*}[htb]
\includegraphics[width=0.88\textwidth]{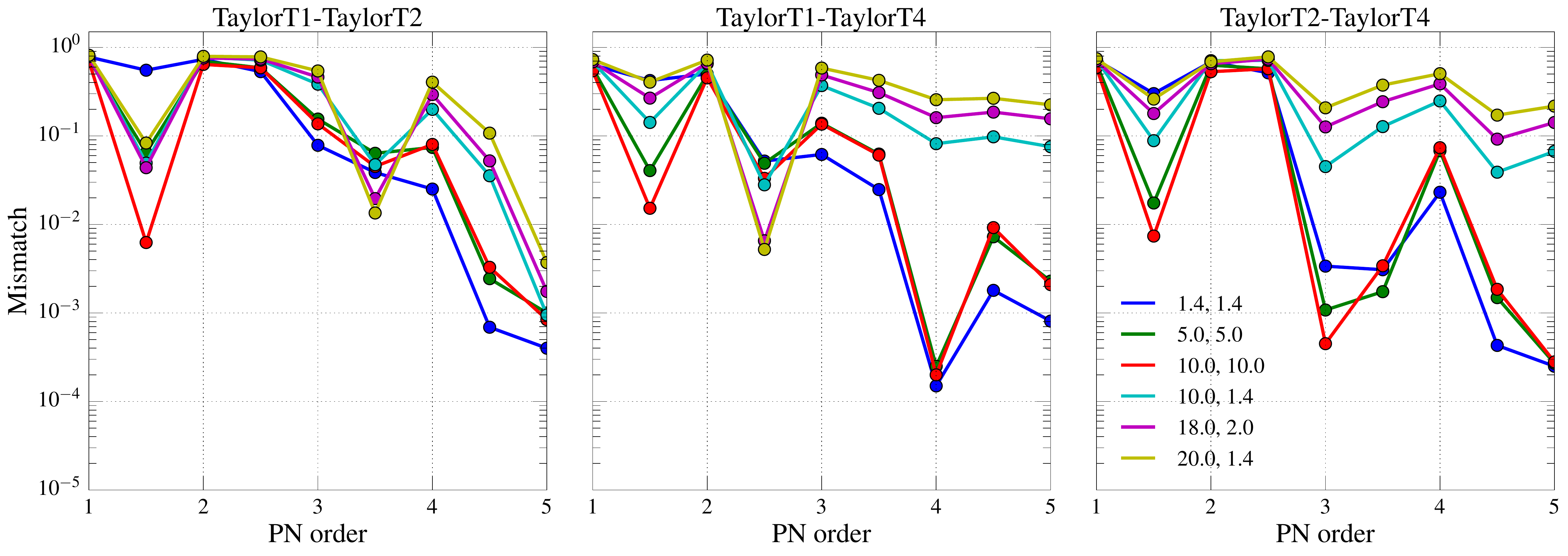}
\caption{Mismatch between approximants as a function PN order. The legend shows the fiducial mass-combinations (in $M_{\odot}$) that we consider. There is a significant reduction in the mismatches at 5pPN (which are consistently below $10^{-2}$), as compared to 3.5PN, between TaylorT1 and TaylorT2. The mismatches involving TaylorT4 also drop for all cases considered.} 
\label{fig:mismatch_approx}
\end{figure*}

Also of interest is the match between approximants. Given that the PN expansions of the binding energy $\mathcal{E}(v)$ and flux $\mathcal{F}(v)$ are known only to a limited PN order, the approximants are not expected to converge towards each other, a fact that becomes manifestly evident during late inspiral. Since our pPN coefficients improve the match of various approximants with EOBNR (often considerably), we thus expect that they will also help the approximants better converge towards each other. Figure~\ref{fig:mismatch_approx} shows the mismatch between different pairs of approximants as a function of the (pseudo) PN order. The figure demonstrates a significant improvement in the agreement between TaylorT1 and TaylorT2. Since the pPN coefficients do not reduce the mismatch between TaylorT4 and EOBNR quite as significantly as for the rest of the approximants, it is not surprising that the effect of the pPN coefficients in ameliorating the difference between TaylorT4 and the rest is not as marked; nevertheless, we still find improvement in the agreement for a good fraction of the mass combinations considered. Note that since TaylorT2 and TaylorF2 are closely related approximants, and given the similarity between the TaylorT2-EOBNR and TaylorF2-EOBNR mismatches, we did not find it necessary to include mismatches involving TaylorF2 in the plots. As expected, we found the latter to be qualitatively similar to mismatches involving TaylorT2.

\subsection{Understanding the results}

As discussed in Appendix~\ref{approx_conv}, it is not guaranteed that pPN coefficients that improve one approximant will necessarily improve another one, since different approximants (e.g., TaylorT2 and TaylorT4) are expanding quantities with different analytic structure. In particular, poles in one case become zeros in the other, so the expansions can have, in principle, very different radii of convergence. However, in Appendix~\ref{approx_conv}, we also obtained sufficient conditions for the pPN coefficients in the flux that improve TaylorT2 to also improve TaylorT1, which is easier to ensure than requiring them to also improve TaylorT4, since the pPN version of TaylorT1 can be thought of as a rational (including logarithms) representation of the pPN version of TaylorT2 (similar to a Pad{\'e} approximant). Here we saw that having small pPN terms at each order (as we obtain with the current method, but had difficulty obtaining with the methods that we previously tried) is sufficient for this to be the case. Indeed, we find that if we extract the pPN flux coefficients through 5pPN from the TaylorT2 $\theta$ coefficients, setting the 5pPN energy coefficients to zero, then these pPN flux coefficients already improve the agreement between TaylorT1 and EOBNRv2. Since we then go on to fit for the 5pPN energy coefficients in TaylorT1, it is not surprising that the final 5pPN TaylorT1 has excellent agreement with EOBNRv2.

\section{Conclusions and Outlook}
\label{sec:conclusions}
Accurate models of GWs from inspiralling compact binaries are a crucial component in the searches for these signals. Given that the construction of template banks will require millions of waveforms to adequately cover regions of the binaries' parameter space, minimizing computational costs is an important challenge. Numerical relativity is able (in principle) to supply accurate waveforms that give the predictions of general relativity. However, such waveforms are extremely expensive to generate, particularly considering the many cycles necessary to completely cover the band of a GW detector~\cite{Szilagyi:2015rwa}. EOBNR waveforms, which calibrate an analytical model to numerical relativity, reproduce the numerical relativity results quite well over large portions of the parameter space and can be generated in a small fraction of the time it takes to produce a numerical relativity waveform. However, they still require solving a nontrivial system of ordinary differential equations, and thus are still too expensive to use to construct template banks directly in most cases.
On the other hand, PN waveforms are quite inexpensive to generate. However, they are unreliable during late inspiral where they begin to diverge from each other as well as from their more accurate NR or EOBNR counterparts. To ameliorate this issue in a manner that does not significantly increase computational costs, we introduce {\it effective} higher order PN terms (which we call ``pseudo-PN'' terms) in the energy and flux functions of the binary designed to improve the agreement between EOBNR (our fiducial waveform of choice) and multiple PN approximants.

We use a procedure described in Sec.~\ref{sec:Intro-pPN}, which, in essence, fits a PN-like ansatz for $t(v)$ and $dv/dt$ to the corresponding quantities computed from EOBNR, to estimate the effective higher order terms in the energy and flux. We compute coefficients for an evenly spaced set of symmetric mass ratios $\eta \in [0.05, 0.25]$, and find that the pPN coefficients seem to vary as low order polynomials in $\eta$ (Fig.~\ref{fig:pPN_eta_fit}). To evaluate the performance of the estimated $\eta$-dependent pPN coefficients, we compare the mismatches of approximants TaylorT1, TaylorT2, TaylorT4, and TaylorF2 evaluated at 5pPN order with our fiducial ``exact'' waveform EOBNR (see Figs.~\ref{fig:mismatch_EOB} and \ref{fig:contour_mismatch}). We find that for approximants TaylorT1, TaylorT2, and TaylorF2, the mismatches with EOBNR over mass-space $m_{1,2} \in [1.4M_{\odot}, 15M_{\odot}]$ are not only smaller than those at 3.5PN, but are around or below $10^{-2}$, often touching $10^{-3}$. The pPN coefficients do not \emph{significantly} improve the performance of TaylorT4, but they do nonetheless reduce the mismatch for high mass ratios and maintain a mismatch of close to $10^{-2}$ for a significant portion of the parameter space considered. Mismatches between approximants also indicate a marked improvement at 5pPN, as compared to 3.5PN, for most of the cases considered. Furthermore, a large fraction of these yield mismatches of below $10^{-2}$ at 5pPN (Fig.~\ref{fig:mismatch_approx}).

Based on the encouraging results produced by our pPN coefficients in boosting the performance of the approximants we have considered, we anticipate similar enhancements in other approximants closely related to ours (see the discussion in Sec.~\ref{sec:adiabatic_pn}). We believe therefore that the pPN coefficients will afford some flexibility in the choice of approximants, were they to be used as part of the pipelines for the detection and parameter estimation of GWs from non-spinning sources.

Of course, the most useful application of our method for GW data analysis would be to extend to the case of spinning binaries. We do not foresee this to be particularly challenging in principle (at least in the case of binaries with non-precessing spins), and plan to pursue this as an extension of the work presented here. In addition, as successful as EOBNR is at reproducing NR waveforms, it may not always parallel full-scale NR in terms of accuracy and reliability. This is particularly the case when spinning systems are considered \cite{Kumar2015,Khan:2015jqa}. Computing pPN coefficients with NR waveforms may therefore be a worthwhile endeavor, given that a number of  long and accurate NR waveforms spanning the full parameter space of interest are becoming available~(see, e.g.,~\cite{SXS-Catalog}). 

\acknowledgments 
We thank K.~G.~Arun, Luc Blanchet, Marc Favata, Mark Hannam, Sascha Husa, Bala Iyer, Daniel Kennefick, Badri Krishnan, Chandra Kant Mishra, and B.~S.~Sathyaprakash for useful discussions. SJK is grateful to the International Centre for Theoretical Sciences (ICTS) for providing hospitality and travel support; he also acknowledges financial support from the Raymond Hughes Foundation at the University of Arkansas. NKJ-M and PA acknowledge support from the AIRBUS Group Corporate Foundation through a chair in ``Mathematics of Complex Systems'' at ICTS. PA's research was, in addition, supported by a Ramanujan Fellowship from the Science and Engineering Research Board (SERB), India, the SERB FastTrack fellowship SR/FTP/PS-191/2012, and by the Max Planck Society and the Department of Science and Technology, India through a Max Planck Partner Group at ICTS. Computations were performed using the ICTS computing clusters Mowgli and Dogmatix. This paper has the LIGO document number LIGO-P1500166-v3.

\appendix

\section{Bounds on the residual for $\varphi(v)$ from the residual for $t(v)$}
\label{phi_bound}

Here we wish to show that, if we minimize the residual between the pPN and EOBNR versions of $t(v)$, it automatically guarantees a small residual for $\varphi(v)$. We will first measure the residual using its maximum, i.e., using the $L^\infty$ norm $\inorm{f} := \max_{v\in[0,v_\text{max}]}|f(v)|$ for $f:[0,v_\text{max}]\to\R$, where we can replace the usual supremum by a maximum since $f$'s domain is compact (and do not have to consider the
essential supremum since we are only applying it to continuous functions). Later we will also use the $L^p$ norms for $p\in\{1,2\}$, defined for a general $p\in[1,\infty[$ by $\pnorm{f} := \left[\int_0^{v_\text{max}}|f(v)|^pdv\right]^{1/p}$ (we will use a lower bound of $v = v_\text{min}$ in the discussion of the matches). The properties of these norms, including the standard inequalities we use here, are discussed in most books on real and/or functional analysis (e.g.,~\cite{Stein2011}).

\subsection{$L^\infty$ bounds}
\label{Linfty_bounds}

We now assume that we have found $t_\text{pPN}(v)$ such that $\inorm{\Delta t}$ is small (where $\Delta$ denotes the difference of the pPN and EOB quantities), and we want to bound $\inorm{\Delta\varphi}$, where $\varphi'(v) = (v^3/M)t'(v)$ [from Eq.~\eqref{energybalance}]. We can thus integrate directly and write
\<
M\inorm{\Delta\varphi} = \max_{v\in[0,v_\text{max}]}\left|\int_0^v \bar{v}^3\Delta t'(\bar{v}) d\bar{v}\right|.
\?
We now integrate by parts to remove the derivative on $\Delta t$ and then apply the triangle inequality (and triangle inequality for integrals) to obtain
\<\label{inf_ibp}
M\inorm{\Delta\varphi} \leq \max_{v\in[0,v_\text{max}]}\Biggl[v^3\left|\Delta t(v)\right| +  3\int_0^v \bar{v}^2\left|\Delta t(\bar{v})\right| d\bar{v}\Biggr].
\?
We are able to discard the boundary term at $v = 0$ in the integration by parts: While $t(v)$ diverges as $v^{-8}$ as $v\searrow0$, both the pPN and EOB $t(v)$s reduce to the PN $t(v)$ for small $v$ and the leading $4$pPN term we are adding is $O[\ln^2(v)]$ for $v\searrow0$, so we know that $v^3\Delta t(v)$ vanishes at least as fast as $v^3\ln^2(v)$ as $v\searrow0$.
We now note that we can bound the maximum of the sum by the sum of the maxima of each term, and each of these maxima can be bounded from above by $v_\text{max}^3\inorm{\Delta t}$ separately. For the first term, this is trivial, and for the second, almost so, after noting that one can bound the
integral from above by its value with $\left|\Delta t(\bar{v})\right| \to \inorm{\Delta t}$, where the latter is a constant we can pull out of the integral (or, alternatively, applying 
the H{\"o}lder inequality in the form $\|fg\|_1 \leq \|f\|_1\inorm{g}$).
Our final result is thus (recalling that we take $v_\text{max} = v_\text{ISCO} = 6^{-1/2}$)
\<
M\inorm{\Delta\varphi} \leq 2v_\text{max}^3\inorm{\Delta t} = \frac{2}{6^{3/2}}\inorm{\Delta t} \simeq 0.136\inorm{\Delta t}.
\?
This inequality is surely not optimal, though it certainly suffices for our purposes. Note that we can replace the lower bound of $v = 0$ by a lower bound of $v = v_0 > 0$, as we use in our fits, with no change, since we
perform the fits setting $\Delta t(v_0) = 0$.

\subsection{$L^2$ bounds}
\label{L2_bounds}

We are also interested in bounds involving the $L^2$ norm, since these are related more directly to matches, as mentioned in Sec.~\ref{sec:pPN_coeffs}. However, going from the $\Delta t$ and $\Delta\varphi$ which we have been considering to bounds on anything involving the waveform is not entirely straightforward, since one has to invert $t(v)$ to obtain the waveform in the time domain. (We could work in the frequency domain, using the stationary phase approximation, but prefer to work in the time domain, since we consider the time domain versions of most approximants.) Nevertheless, if we take $\Delta t = 0$, so we just consider
$\Delta\varphi$, and also take the amplitudes of the two waveforms to be equal, then it is in fact straightforward to give a lower bound on the white noise match in terms of $\tnorm{\Delta\varphi}$, if we further assume that we have a long enough observing time that we are able to discard rapidly varying contributions to the integrals.

We thus first note that the white noise match between two waveforms $h_1$ and $h_2$ is given by 
\<
\label{cMw}
\cM_w(h_1,h_2) := \max_{t_c,\varphi_c}\frac{\tip{h_1}{h_2}}{\tnorm{h_1}\tnorm{h_2}}\geq \frac{\tip{h_1}{h_2}}{\tnorm{h_1}\tnorm{h_2}},
\?
where $t_c$ and $\varphi_c$ denote the time and phase at coalescence, respectively, and $\tip{f}{g} := \int_{t_\text{min}}^{t_\text{max}} f(t)g(t)dt$ is the standard $L^2$ inner product for real functions on the interval $[t_\text{min},t_\text{max}]$. Now, we can write the waveforms as $h_k(t) = A(v(t))\cos\varphi_k(v(t))$, where $v(t)$ is the same for both waveforms, by assumption, so $\tnorm{h_k} \simeq \tnorm{A\circ v}/\sqrt{2}$, where we have written $\cos^2\varphi_k = (\cos 2\varphi_k + 1)/2$ and neglected the contribution of the oscillatory $\cos 2\varphi_k$ piece. If we now write $\Delta\varphi := \varphi_1 - \varphi_2$, so $\cos\varphi_1\cos\varphi_2 = [\cos\Delta\varphi + \cos(\varphi_1 + \varphi_2)]/2$, and again neglect the oscillatory $\cos(\varphi_1 + \varphi_2)$ term, we have
\<
\begin{split}
\tip{h_1}{h_2} &\simeq \frac{1}{2}\int_{t_\text{min}}^{t_\text{max}}A^2(v(t))\cos\Delta\varphi(v(t))dt\\
&= \frac{1}{2}\int_{v_\text{min}}^{v_\text{max}}A^2(v)\cos\Delta\varphi(v)t'(v)dv.
\end{split}
\?
Now, $\cos x \geq 1 - x^2/2$ (which can be obtained by antidifferentiating $\cos x \leq 1$ twice), so we have
\<
\cM_w(h_1,h_2) \geq 1 - \frac{\inorm{A^2t'}\tnorm{\Delta\varphi}}{2\tnorm{A\circ v}^2},
\?
where we have used the inequality $\|fg^2\|_1 \leq\inorm{f}\tnorm{g}$, which can be obtained directly or from the H{\"o}lder inequality.
Here $\inorm{A^2t'}$ and $\tnorm{A\circ v}$ are both finite, since we have $v_\text{min} > 0$, so in this simple case, we can see that bounding $\tnorm{\Delta\varphi}$
to be close to zero implies that the matches should be close to unity. Note that we will consider norms over the full interval $[0,v_\text{max}]$ in the subsequent discussion, but
the norms over the smaller interval we consider here are, of course, bounded from above by those over the larger interval.

Thus, we note that we have
$\tnorm{\Delta\varphi}\leq v_\text{max}^{1/2}\inorm{\Delta\varphi} \simeq 0.64\inorm{\Delta\varphi}$.
We can also obtain a bound on $\tnorm{\Delta\varphi}$ in terms of $\tnorm{\Delta t}$. Here we start from the same
integrated-by-parts expression for $\Delta\varphi$ we used above to obtain Eq.~\eqref{inf_ibp} to now obtain
\<
\begin{split}
M^2\tnorm{\Delta\varphi}^2 &= \int_0^{v_\text{max}}\Biggl\{v^6\left[\Delta t(v)\right]^2 - 6v^3\Delta t(v)\int_0^v\bar{v}^2\Delta t(\bar{v})d\bar{v}\\
&\quad + 9\left[\int_0^v\bar{v}^2\Delta t(\bar{v})d\bar{v}\right]^2\Biggr\}dv\\
&\leq v_\text{max}^6\tnorm{\Delta t}^2 + \int_0^{v_\text{max}}\left[6v^3|\Delta t(v)|\,\tnorm{s}\tnorm{\Delta t} + 9\tnorm{s}^2\tnorm{\Delta t}^2\right]\\
&\quad\times dv\\
&\leq \left(v_\text{max}^6 +6\tnorm{c}\tnorm{s} + 9v_\text{max}\tnorm{s}^2\right)\tnorm{\Delta t}^2.
\end{split}
\?
Here we have defined $c(v) := v^3$ and $s(v) := v^2$ (so $\tnorm{c} = v_\text{max}^{7/2}/\sqrt{7}$ and $\tnorm{s} = v_\text{max}^{5/2}/\sqrt{5}$) and again used the triangle inequality (and triangle inequality for integrals) to obtain the first inequality, along with the Cauchy-Schwarz version of the H{\"o}lder inequality, $\|fg\|_1\leq\tnorm{f}\tnorm{g}$, from which we obtain
\<
\label{int_bound}
\int_0^v|s(\bar{v})\Delta t(\bar{v})|d\bar{v} \leq \int_0^{v_\text{max}}|s(\bar{v})\Delta t(\bar{v})|d\bar{v} = \|s\Delta t\|_1 \leq \tnorm{s}\tnorm{\Delta t}
\?
(we also use the version with $s\to c$).
We thus have
\<
M\tnorm{\Delta\varphi} \leq \sqrt{\frac{14}{5} + \frac{6}{\sqrt{35}}}v_\text{max}^3\tnorm{\Delta t} \simeq 0.133\tnorm{\Delta t}.
\?
In these $L^2$ inequalities, replacing the lower bound of $v = 0$ by $v = v_0 > 0$ (as we actually do in the fits) would at most just reduce the size of the constants in the final inequalities: The only change to the derivation is the different
integration interval, since the basic integrated by parts expression does not change in this case, as discussed above.

\section{
Prospects for improving more than one approximant with the same pPN coefficients}
\label{approx_conv}

As a first example, we consider a function of the form $\cR(v) := \cF(v)/\cE'(v)$ for $v\in[0,v_\text{max}]$, where we know $\cF$ and $\cE$ to some approximation. We want to know whether improving the accuracy of $\cR$ by adding additional terms to $\cF(v)$ and $\cE(v)$ will necessarily also improve the accuracy of other approximants to $\cR$, consisting of $T_n[\cR]$ and $1/T_n[1/\cR]$, where $T_n[\cdot]$ denotes the $n$th order Maclaurin expansion (i.e., Taylor expansion about $0$) of its argument. We thus note that if $\cF(v) := A v^a[1 + \ff(v)]$ and $\cE'(v) := B v^b[1 + \fe(v)]$ with $A,B\in\R$ and $a, b\in\N$ constants, $\fe$ and $\ff$ real analytic, and $|\fe(v)|, |\ff(v)| < 1$ for $v\in[0,v_\text{max}]$, then one can obtain an arbitrarily good approximation to $\cR$ from any of these approximants by using sufficiently high-order Maclaurin expansions of $\cF$ and $\cE$ (and sufficiently large $n$) since the Maclaurin series for both $\cF$ and $\cE$ converge to the functions and the series coming from expanding $1/\cE'$ or $1/\cF$ also converge, due to the assumption about the absolute values of $\fe$ and $\ff$.

Of course, this is not the situation we are actually in. For instance, we are actually first fitting to $T_n[1/\cR]$, not $\cR$. We also know that we have $\ln^n(v)$ terms in the expansion of both the energy and flux and that they both diverge at the light ring in the extreme mass-ratio limit (where $v = 3^{-1/2} \simeq 0.577$), as discussed in, e.g.,~\cite{DIS01}. However, since we are considering $v_\text{max} = v_\text{ISCO}$, this divergence does not necessarily concern us, and we in fact find that the true energy flux (as obtained from NR, EOBNR, or black hole perturbation theory) is well within a factor of $2$ of the Newtonian energy and flux over this range (see, e.g., Fig.~4 in~\cite{Damour2009} in the equal-mass case and Fig.~1 in \cite{DIN} in the extreme mass-ratio limit), so that we have $|\ff(v)| < 1$ there. The binding energy in the test mass limit is also well-behaved up to $v_\text{ISCO}$, and we also have
$|\fe(v)| < 1$ for $v \leq v_\text{ISCO}$ in that limit. (There does not appear to be any numerical relativity data for the binding energy versus GW frequency for comparable-mass binaries against which to compare in the literature.)

Perhaps more importantly, we are not considering making Maclaurin expansions to arbitrary order: We are rather adding on a small number of additional terms to a known series (about $v = 0$, but not exactly a Maclaurin series due to the logarithms) and taking $n$ in the other approximants $T_n[\cR]$ and $1/T_n[1/\cR]$ to be consistent with the added terms. We are obtaining the first six additional terms by (basically) minimizing the $L^2$ norm of the difference between the true value of $1/\cR$ and our $T_n[1/\cR]$. We then fix the final two additional terms by minimizing the difference between the true value of $\cR$ and our (unexpanded) $\cR$.

Actually, we minimize the antiderivatives of $1/\cR$ and $T_n[1/\cR]$ for the first fits (the ones involving TaylorT2), but we ignore this slight complication here, since we are only interested in the accuracy of the antiderivatives themselves, as they are what gives the waveform, and controlling the function itself with an $L^p$ norm is sufficient to control the antiderivative, since we are working on a compact set [cf.\ Eq.~\eqref{int_bound}]. Thus, our assumption about the smallness of the norm of $1/\cR - T_n[1/\cR]$ is sufficient to guarantee that the norm of the antiderivatives will also be small. However, we cannot claim \emph{a priori} that the fit of the antiderivatives will give us good agreement for the function itself (though this will likely be the case).

As an example of the situations under which we can expect to get an improvement of TaylorT1 by fitting to improve TaylorT2, consider the case where we are just considering ppN terms in the energy flux (which is the case at $4$ and $4.5$pPN, and can be taken to be the case at $5$pPN when we are just looking at TaylorT2), so we are first fitting $T_N[\cE'_\text{known}/\cF_{\leq n\text{pPN}}]$ to $\cE'_\text{EOB}/\cF_\text{EOB}$ and want to know if this will improve the fit of $\cF_{\leq n\text{pPN}}/\cE'_\text{known}$ to $\cF_\text{EOB}/\cE'_\text{EOB}$. We thus write $\cF_{\leq n\text{pPN}} = \cF_\text{known} - \cF_{n\text{pPN}}$ (where we introduce the minus sign to slightly simplify things later) and take $N = 2n$.\footnote{Here we use the subscript $\leq n\text{pPN}$ to denote that we are including all the terms up to a given pPN order, which we just denoted by $n\text{pPN}$ previously, since we now reserve that subscript for just the $n$pPN terms.} We thus want to bound $\pnorm{\cF_{\leq n\text{pPN}}/\cE'_\text{known} - \cF_\text{EOB}/\cE'_\text{EOB}}$ in terms of $\pnorm{T_{2n}[\hat{\cE}'_\text{known}/\hat{\cF}_{\leq n\text{pPN}}] - \hat{\cE}'_\text{EOB}/\hat{\cF}_\text{EOB}}$, $\pnorm{T_{2n}[\hat{\cE}'_\text{known}/\hat{\cF}_\text{known}] - \hat{\cE}'_\text{known}/\hat{\cF}_\text{known}}$, and $\pnorm{\cF_{n\text{pPN}}}$, in addition to known constants, where hats denote scaling by the Newtonian values (after taking the derivative, for $\cE'$) and we consider an arbitrary $L^p$ norm with $p\in[1,\infty]$. The
Newtonian scaling is necessary for the first two norms to be finite, since $\cE'/\cF$ goes as $v^{-9}$ as $v\searrow 0$.

We first note that we can write
\begin{equation}
T_{2n}\left[\frac{\hat{\cE}'_\text{known}}{\hat{\cF}_{\leq n\text{pPN}}}\right] = T_{2n}\left[\frac{\hat{\cE}'_\text{known}}{\hat{\cF}_\text{known}}\right] + \hat{\cF}_{n\text{pPN}},
\end{equation}
where $\cF_N$ and $\cE_N$ are the Newtonian flux and binding energy, respectively.
We thus have (adding a convenient zero)
\<
\begin{split}
&T_{2n}\left[\frac{\hat{\cE}'_\text{known}}{\hat{\cF}_{\leq n\text{pPN}}}\right] - \frac{\hat{\cE}'_\text{EOB}}{\hat{\cF}_\text{EOB}}\\
& = T_{2n}\left[\frac{\hat{\cE}'_\text{known}}{\hat{\cF}_\text{known}}\right]  - \frac{\hat{\cE}'_\text{known}}{\hat{\cF}_\text{known}} + \hat{\cF}_{n\text{pPN}} + \frac{\hat{\cE}'_\text{known}}{\hat{\cF}_\text{known}} - \frac{\hat{\cE}'_\text{EOB}}{\hat{\cF}_\text{EOB}}.
\end{split}
\?
Now, we can write the last three terms as (adding another convenient zero)
\<
\frac{\hat{\cE}'_\text{known}\hat{\cE}'_\text{EOB}}{\hat{\cF}_\text{known}\hat{\cF}_\text{EOB}}\left[\left(\frac{\hat{\cF}_\text{known}\hat{\cF}_\text{EOB}}{\hat{\cE}'_\text{EOB}} - 1\right)\frac{\hat{\cF}_{n\text{pPN}}}{\hat{\cE}'_\text{known}}+ \frac{\hat{\cF}_\text{EOB}}{\hat{\cE}'_\text{EOB}} - \frac{\hat{\cF}_{\leq n\text{pPN}}}{\hat{\cE}'_\text{known}}\right].
\?
Thus, applying the reverse triangle inequality twice, and noting that $\pnorm{f/g}\geq\pnorm{f}/\inorm{g}$ (which is just a rewriting of a case of the H{\"o}lder inequality used earlier for $p < \infty$ and trivially true for $p=\infty$), we find that
\<
\begin{split}
&\pnorml{T_{2n}\left[\frac{\hat{\cE}'_\text{known}}{\hat{\cF}_{\leq n\text{pPN}}}\right] - \frac{\hat{\cE}'_\text{EOB}}{\hat{\cF}_\text{EOB}}}\\
&\geq \left[\pnorml{ \frac{\cF_\text{EOB}}{\cE'_\text{EOB}} - \frac{\cF_{\leq n\text{pPN}}}{\cE'_\text{known}}}- \pnorml{\left(\frac{\hat{\cF}_\text{known}\hat{\cF}_\text{EOB}}{\hat{\cE}'_\text{EOB}} - 1\right)\frac{\cF_{n\text{pPN}}}{\cE'_\text{known}}}\right]\\
&\quad \times \inorml{\frac{\hat{\cF}_\text{known}\cF_\text{EOB}}{\hat{\cE}'_\text{known}\cE'_\text{EOB}}}^{-1} - \pnorml{ T_{2n}\left[\frac{\hat{\cE}'_\text{known}}{\hat{\cF}_\text{known}}\right] - \frac{\hat{\cE}'_\text{known}}{\hat{\cF}_\text{known}}}.
\end{split}
\?
Therefore,
\<\label{T2toT1bound}
\begin{split}
&\pnorml{\frac{\cF_{\leq n\text{pPN}}}{\cE'_\text{known}} -  \frac{\cF_\text{EOB}}{\cE'_\text{EOB}}}\\
&\leq \left\{\pnorml{T_{2n}\left[\frac{\hat{\cE}'_\text{known}}{\hat{\cF}_{\leq n\text{pPN}}}\right] - \frac{\hat{\cE}'_\text{EOB}}{\hat{\cF}_\text{EOB}}} + \pnorml{ T_{2n}\left[\frac{\hat{\cE}'_\text{known}}{\hat{\cF}_\text{known}}\right] - \frac{\hat{\cE}'_\text{known}}{\hat{\cF}_\text{known}}}\right\}\\
&\quad\times \inorml{\frac{\hat{\cF}_\text{known}\cF_\text{EOB}}{\hat{\cE}'_\text{known}\cE'_\text{EOB}}} + \pnorml{\left(\frac{\hat{\cF}_\text{known}\hat{\cF}_\text{EOB}}{\hat{\cE}'_\text{EOB}} - 1\right)\frac{\cF_{n\text{pPN}}}{\cE'_\text{known}}}.
\end{split}
\?

This inequality seemingly implies that in order for us to have our fit for TaylorT2 also improve TaylorT1 (or TaylorT4), we should also have the higher PN expansion of $\hat{\cE}'_\text{known}/\hat{\cF}_\text{known}$
agreeing well with the unexpanded $\hat{\cE}'_\text{known}/\hat{\cF}_\text{known}$ and also a small $\cF_{n\text{pPN}}$ (and we indeed saw that smaller pPN coefficients helped the TaylorT2 fits to also improve the
agreement of TaylorT1 and---to a lesser extent---TaylorT4). However, this is just an upper bound, and is likely not sharp, particularly since we have used the weaker version of the reverse triangle inequality without
the absolute value (i.e., the first and last terms in $\|x - y\|\geq\bigl|\|x\| - \|y\|\bigr|\geq \|x\| - \|y\|$ instead of the first and second terms). Indeed, at $4$pPN, the right-hand side of Eq.~\eqref{T2toT1bound} is $\sim 50$--$100$ times larger than the left-hand side if we take $p = 1$, $2$, or $\infty$. The difference between the two sides is not so pronounced ($\lesssim 50$) if we consider the case where we include the Newtonian scaling on all terms, where the inequality also holds. Thus, one cannot use this equality to say that a small $\cF_{n\text{pPN}}$ is necessary for the
TaylorT2 fit to also improve TaylorT1, though this is certainly a sufficient condition (along with having appropriately small values for the other two norms). Indeed, in the $4$pPN examples mentioned above, the contribution from the final term is at most a quarter of the total, and often much less.

\bibliography{PseudoPN}

\end{document}